# A History of the Magellanic Clouds and the European Exploration of the Southern Hemisphere

Michel Dennefeld[1]

[1] Institut d'Astrophysique de Paris (IAP), CNRS & Sorbonne Université, France

**The Magellanic Clouds were known before Magellan's voyage exactly 500 years ago, and were not given that name by Magellan himself or his chronicler Antonio Pigafetta. They were, of course, already known by local populations in South America, such as the Mapuche and Tupi-Guaranis. The Portuguese called them Clouds of the Cape, and scientific circles had long used the name of Nubecula Minor and Major. We trace how and when the name Magellanic Clouds came into common usage by following the history of exploration of the southern hemisphere and the southern sky by European explorers. While the name of Magellan was quickly associated to the Strait he discovered (within about 20 years only), the Clouds got their final scientific name only at the end of the 19th century, when scientists finally abandoned Latin as their communication language.**

This year we celebrate the 500th anniversary of the discovery of the navigable sea route that separates mainland South America from Tierra del Fuego — now known as the Strait of Magellan — by Fernão de Magalhães (Ferdinand Magellan in English) and his companions. It therefore seems an appropriate time to examine the "history" of the Magellanic Clouds, not least because the study of the Clouds was one of the main reasons for the foundation of astronomical observatories in Chile.

Magellan's expedition entered the strait at Cabo de las Virgenes on 21 October 1520 and exited via Cabo Deseado on 28 November. The Clouds are mentioned right after they left the strait, in the best known narrative of the voyage — that by Antonio Pigafetta, who writes:

*Il polo antartico non ha stella alcuna della sorte del polo artico, ma si veggon molte stelle congregate insieme, che sono come due nebule* [the Clouds], *un poco separate l'una dall'altra, e un poco oscure nel mezzo. Tra queste ne sono due, non molto grandi né molto lucenti, che poco si muovono: e quelle due sono il polo antartico.* (see Ramusio, 1550).

Obviously neither Pigafetta nor Magellan himself named them the "Magellanic Clouds" — nor indeed did they name the strait after Magellan! The discovery of the passage between the Mar del Oceano (the Atlantic) and the Mar del Sur (the Pacific) quickly became known to sailors and navigators venturing to these remote countries. The race to reach the Spice Islands (the Moluccas) and the competition between Portugal, which had exclusivity of the "eastern route" (via the Cape of Good Hope) since the Treaty of Tordesillas (1494) and Spain which wanted to exploit the "western route" to take possession of new lands, lent great importance to the discovery. Before long this new strait came to be called the “Strait of Magellan” after its discoverer. Of course, the various water channels at the southern end of South America were well known by local fishermen (Alacalufes or Tehuelches), but unfortunately no written reports by locals seem to exist.

As regards the Clouds, history is less clear. Magellan was not the first to reach southern latitudes where the Clouds are visible. On each sea expedition, at least one pilot/astronomer was present to determine the ship’s position by astronomical means, so the Clouds could not have escaped their view. However, accounts are few. There are likely two reasons for this. First, the navigators were looking primarily for some star or asterism analogous to the "Tramontana" (the North Polar Star) in the south in order to measure their latitude and obviously the Clouds would not serve that purpose. Second, at that time travel documents were kept secret (for instance in the Casa de Contratacion in Sevilla, or in the Casa da India in Lisbon) in view of the competition between the various countries. So, to understand the “history” of the Clouds, one needs to follow the history of discoveries and travel in the south, and also the history of the mapping of the southern sky by European explorers and astronomers.

Local populations in South America of course had a knowledge of the southern sky long before Europeans reached that hemisphere. Unfortunately, little is known about their observations as information was transmitted orally and it is only in recent times that efforts have been made to record some of this knowledge before it is lost forever. For instance, the Tupi-Guaranis, in the region of Rio de Janeiro in Brazil, compare the Clouds to fountains (*Hugua*) where a tapir (in the LMC) or a pig (in the SMC) is drinking (Afonso, 2006). Interestingly, the Mapuche in Chile also compare the Clouds to water ponds, called *Rüganko* or *Menoko* in their local language (e.g. Pozo Menares et al., 2014). There were initially three ponds, but one has already dried out (the Coal Sack ?), a second one is presently on the way to dry out (the SMC) and when the last one (LMC) will also become empty, it will be the end of the Universe! Modern calculations (e.g. Wang et al. 2019) show indeed that a few hundred million years ago, one of the ponds did indeed loose some "water" to create the Magellanic Stream, but we still have one or two billions of years before the final "dry out". These water ponds are in the *Wenu Mapu*, the heavens above, and are associated with the *Wenu Leufu*, the river above, i.e., our Milky Way . The similar nature of the Milky Way and the Clouds was therefore recognised long ago.

## Early explorers

Little is known about travel to the south by the Greeks or the Egyptians, although the latter traveled down the Red Sea to seek incense and gold. The first systematic travel in this direction is by the Arabs, progressing along the east coast of Africa or directly to India. When the Portuguese and Spaniards reached India and the Moluccas, they found a lot of trading posts already established by Arabs in India, Indonesia and Malaysia. From there spices were traded, shipped to the Arabic peninsula by boat and then moved further inland. They did then reach Europe via the trade monopoly of Venetian or Genovese ships. Several nautical sources are available to us, dating back to the Persians or the Arabs. The best known is the *Kitab al-Fawaïd* ("Book of (nautical) principles") of 1475 by the *mu'allim* (master of navigation) Ahmad ibn Majid: this knowledge was so precious that Vasco de Gama enrolled a "*Mu'allim Canaca*" (literally, "pilot astrologuer") to cross the Indian ocean in 1498 and reach Calicut (today's Kozhikode). It is clear from those treatises that the Arabs navigated by the stars. They used the Pole Star (*Gah*), with its two "Farkad" (litteraly, veals, but better known as "guards", β and γ UMi), and also the big chariot (Ursa Major; *Na's* ) or the Pleiades (*at-Turayyä*) and then further down *as-Suhayl* (Canopus) or other southern stars — the stars of the Southern Cross were included in Centaurus ( Ferrand, 1928). The location of ports, or characteristic features of the coast were always defined in latitude with reference to the height of Polaris or other stars, whose altitudes were defined in" fingers" above horizon, or azymuth by the raising or setting of some specific stars: Khoury (1987) discusses the nautical poems of Ibn Majid, which are a more detailed expansion of his Hāwiya of 1461. The Clouds were definitely seen, named as-Sahā'ib, or al-Ğamāma in those poems; they were called "clouds of the south pole". For instance, in his nautical poem of Sofala, "As-Sufaliyya" (~1465), Ibn Majid says: "*There are two White Clouds, ô brother: one is visible to the naked eye, the other one is faint. The position of the White Clouds is between Canopus and Sirius; but it is at a distance of ten fingers from Canopus, listen to my discourse, that is one arrow; and at a distance of two arrows from Sirius and you can see both of them in straight line with the naked eye*" (translation by I. Khoury (1983). But they were of no use for navigation owing to their diffuse nature: an error of one degree in latitude is equivalent to about 110 km.

The most ancient record we have of sky observations that include a possible mention of the Clouds seems to be the *Suwar al-Kawakib* (Book of Fixed Stars), written around 964 by the Persian astronomer Abd-al-Rahman al-Sufi. Ludwig Ideler, translating it into German from an extract by Kazvini (Ideler, 1809), includes the following statement:

*Unter den Füssen des Suhel* (the classical name for Canopus) *steht, wie einige behaupten, ein weisser Fleck, ... den man in Tehama El-bakar, den Ochsen nennt.*

This is believed to be the Large Cloud, which could be seen in good weather conditions from the south of Arabia, although no coordinates were given. In 1874, a translation into French was made by the Danish astronomer Hans Schjellerup directly from an original manuscript; this version is slightly different :

*Le vulgaire croit qu'il y a au-dessous des pieds du Suhail quelques étoiles luisantes et blanches qui ne se font voir ni dans Irak ni dans Nadschd, et que les habitants du Tihamat nomment ces étoiles al-bakar, les Vaches.*

Schjellerup refers to stars and not a cloud; and *al-bakar* does indeed mean cows or a herd, rather than a bull. The accompanying illustration of the constellation Argo-Navis shows a group of stars, below Suhail (Canopus, indicated by the red arrow in Figure 1, at the southern extremity of the ship's rudder). The illustration on the globe brought back by John Malcolm (Dorn, 1829) shows also a conglomerate of stars south of Canopus rather than a cloud, but the position would be approximately correct. The later catalogue of Ulugh-Beg (1437) does not add on that, as, from his own observations in Samarkand, he admits that he could not recover 27 (southern) stars from the previous Al-Sufi catalogue.

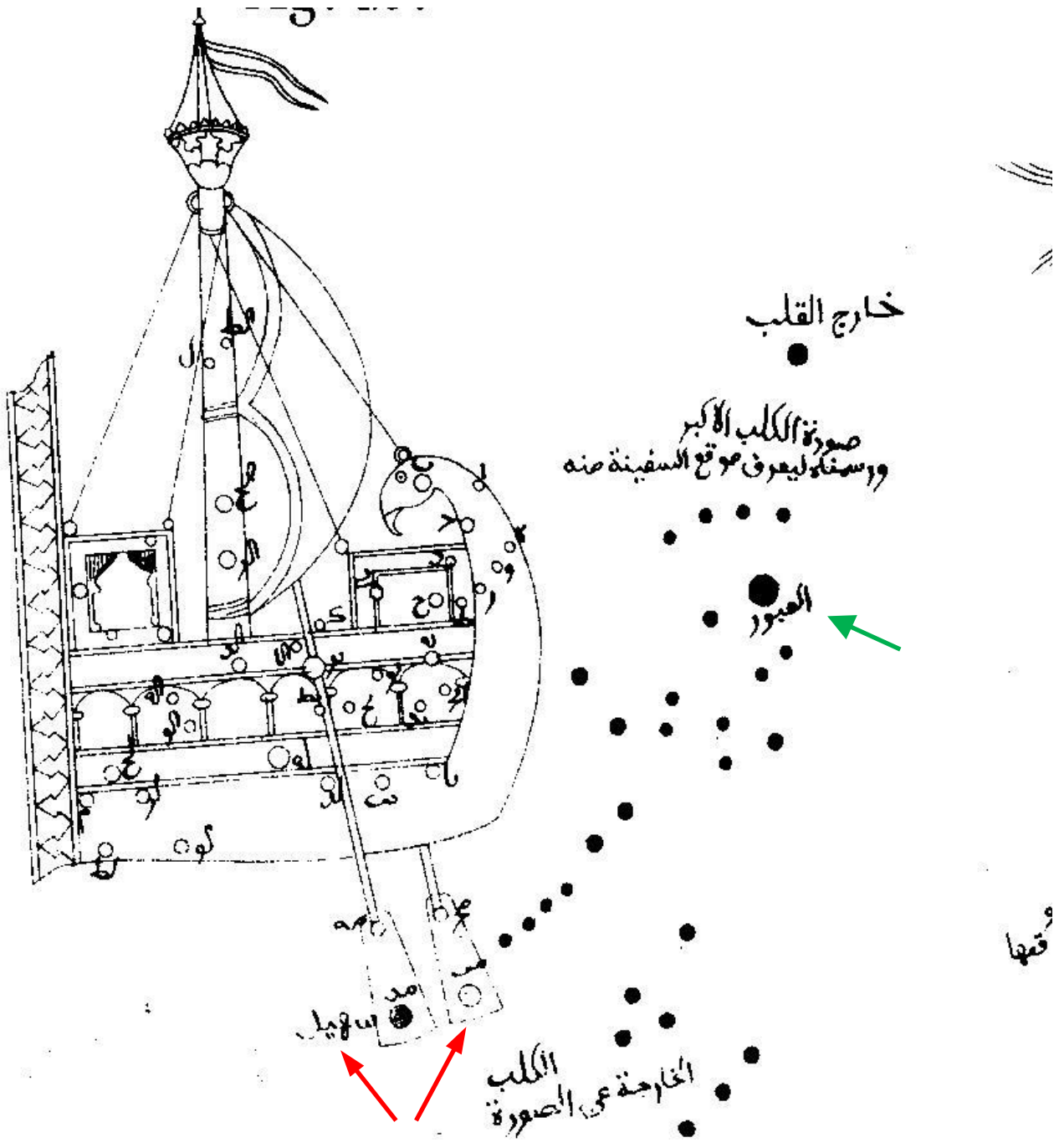

Figure 1 Argo - Al Sufi ( St. Petersburg manuscript, Schjellerup, 1874 ). The words at the bottom say "Dog, outside the image". Red arrows: as-Suhayl (Canopus) Green arrow: Sirius (marked "Al-Obour", Transient !)

The Clouds can be seen below about latitude +15 degrees so Chinese voyagers, having long-established contacts with the Arab world, would have seen them when navigating around India. Admiral Zheng He, for example, made seven voyages to Arabia, beginning in 1405. However, these explorations were stopped by the Xuande Emperor (who reigned 1425–1435), China closing-up again, and orders were even given to destroy all documents relating to the voyages, so we have no records about their observations today (see Levathes, 1994).

Portuguese sailors, progressing systematically south, were probably the first western explorers to see the southern sky. Alvise de Ca'da Mosto 's, during his travel to the river Gambia (about $+13^{\circ}$ N) in summer 1455, mentions the Southern Cross (first known mention, without Christian reference), but not the Clouds, probably because they were not easely seen in July. So the first reference to the Clouds seems to be that by Amerigo Vespucci during his third voyage in 1501–1502:

*E fra le altre viddi tre Canopi: i due erano molto chiari, il terzo era fosco e dissimile dalli altri* (Mundus Novus, 1504).

The two "clear” clouds and the third "dark and different from the others" are now interpreted as meaning the two Clouds and the Coalsack, although no precise coordinates are given. But he sketched southern stars at various times during the night (with some accompanying nebulosity, either white or black, see Fig. 2), and said that he measured their “diameter”, which means the circle they describe around the south pole. This illustrates his skills, as the orientation of the asterism is fundamental to estimate its precise position with respect to the South Pole itself (in the same way as "the guards", β and γ UMi, were used in the north for the Polar star). He also states that he had taken precise measurements of many stars, but had sent his notebook to the king (as proofs of his voyage and skill, presumably) and was waiting to get it back to write a more detailed report about all his observations. This never happened, unfortunately for us, but Vespucci nevertheless was appointed by the King of Spain as the first Piloto Mayor (Chief Pilot) of the newly founded Casa de Contratacion in Sevilla in 1508, a proof that he was highly appreciated. His travel reports were published, under others, in his "Lettere" and in the famous book “Mundus Novus” (1504), which is also at the origin of the name "America" given by the cartographers of the Duke of Lorraine in St Dié (France) to the newly discovered lands on the map which they produced in 1507 (Cosmographiae Introductio). But this is another story…

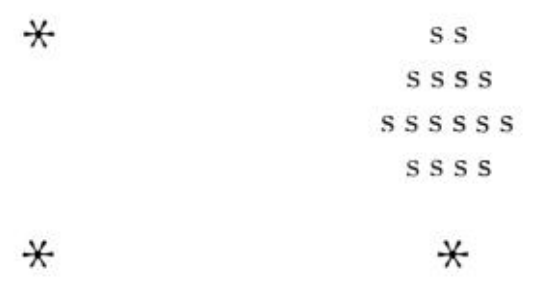

Fig.2 One of Vespucci’s sketches, with a white cloud (in *Mundus Novus*)

A fleet of 13 Portuguese ships, led by Pedro Alvares Cabral and bound for the Moluccas (the “Spice Islands” in Indonesia) to exploit the route opened by Gama, made a significant diversion and reached Brazil (called at that time Tierra de la Vera Cruz) at the latitude of Bahia ($\sim 14^{o}$ S) in April 1500. One of the ships was immediately sent back to Portugal to transmit the news of the discovery[1], carrying with it three letters of report, all dated May 1st. Besides the one by Pero Vaz da Caminha, the official secretary of Cabral, considered as the official reconnaissance of Brazil, the one by Mestre João (João Faras), physician and navigator, is of most interest to us: it contains a sketch of the southern sky (see Figure 3a) where the Southern Cross can easily be identified, with the pointers α and β Cen called “guardas”, as well as some other stars like the southern triangle. The term "la bosya", in the lower left, is the one used in the north to refer to the small chariot, Ursa Minor, in reference to the Pole Star. This clearly shows that, at least at the beginning, the aim was always to find an asterism able to represent the south pole to aid navigation; this was later abandoned in favour of using the altitude of the Sun to derive latitudes. But there is no mention of the Clouds here (the sketch shows that they would fall at the edge of the map), probably because they are of no use for navigation and were low on the horizon in April-May.

After the Portuguese settled in Kochi (Cochin), Afonso de Albuquerque launched an expedition in 1511 to take Malaca and reach further east to the spice islands. But one has to wait until Andrea Corsali travelled to India and wrote two letters from Kochi, in 1515 and 1517, describing his voyage and experiences, to get further news about the sky. The first letter contains a description of the southern sky (Figure 3b). The various stars are more difficult to recognize than in Faras's letter, but the Southern Cross is there, and the two Clouds are also seen, with a star in the middle which is presumably γ Hydri. He writes:

*In che luogo sia il polo antartico, ....ed evidentemente lo manifestano <u>due nugolette di ragionevol grandezza (</u>*the Clouds*), ch'intorno ad essa continuamente ora abbassandosi e ora alzandosi in moto circulare camminano, con una stella sempre nel mezzo, la qual con esse si volge lontana dal polo circa undici gradi* (Ramusio, 1550)

and he continues about the Southern Cross. This, from 1515, seems to be the first clear representation of the Clouds. But it is not clear when the mentions of the Cross and the Antarctic Pole (whose position on the sketch is closer to the ecliptic pole rather than the equatorial one) were added: Rychard Eden, in his translation (1555) of d'Anghiera's *De Orbe Novo,* has this figure with the names — but some earlier manuscripts do'nt! Several versions of this figure have been published, not always correctly attributed: some even attributed it to Magellan whose travel reports never included any sketch! D'Anghiera, in his compilation of travels (d'Anghiera, 1516, in latin, third decade), says "*the Portuguese have gone beyond the fifty-fifth degree of the other pole, where ... they could see throughout the heavenly vault certain nebulae, similar to the Milky Way, in which rays of light shone" (translation by F. McNutt)*. There is no mention of where he got this from but the description resembles Vespucci's.

| | |
|---|---|
| 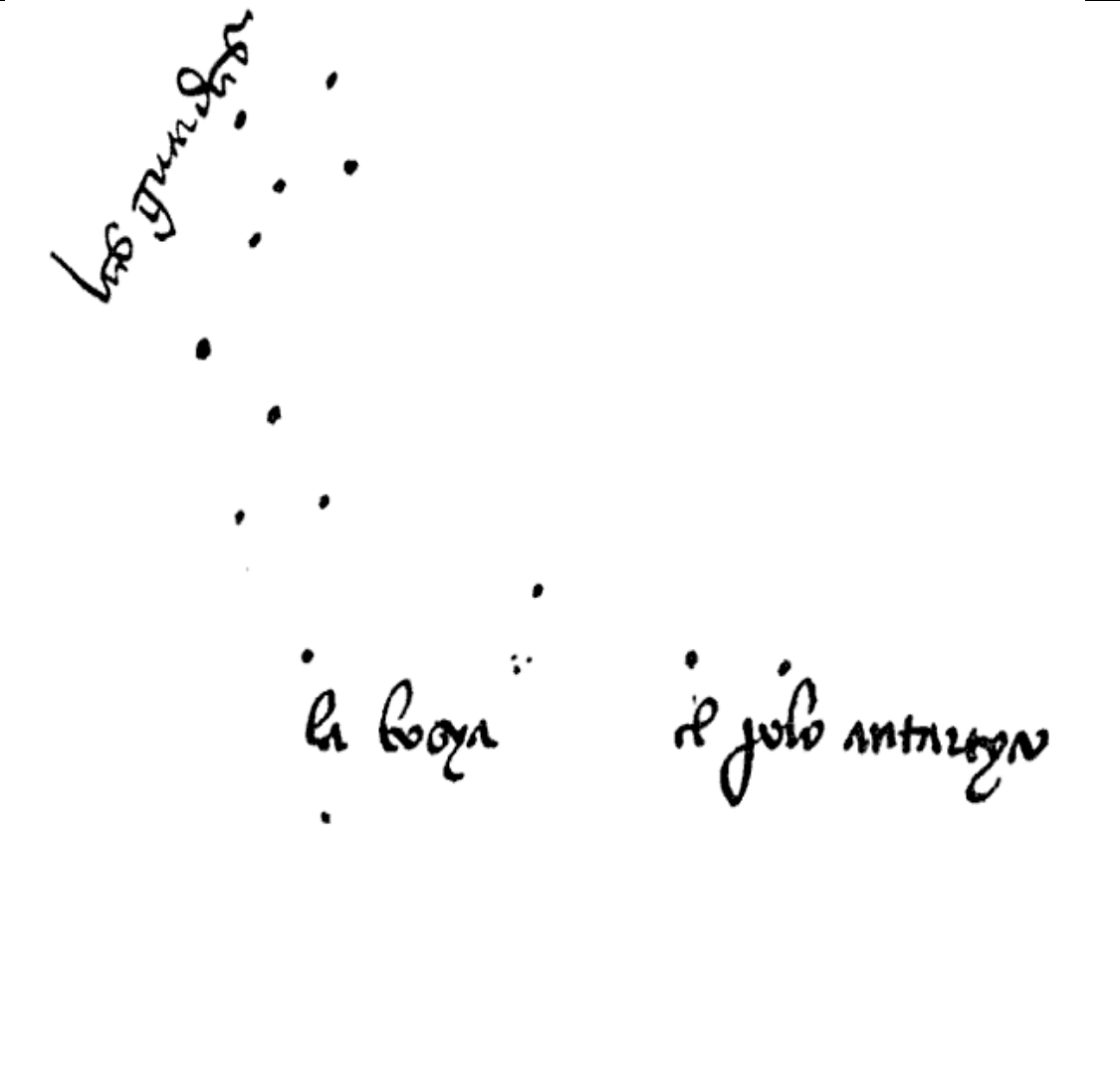 | 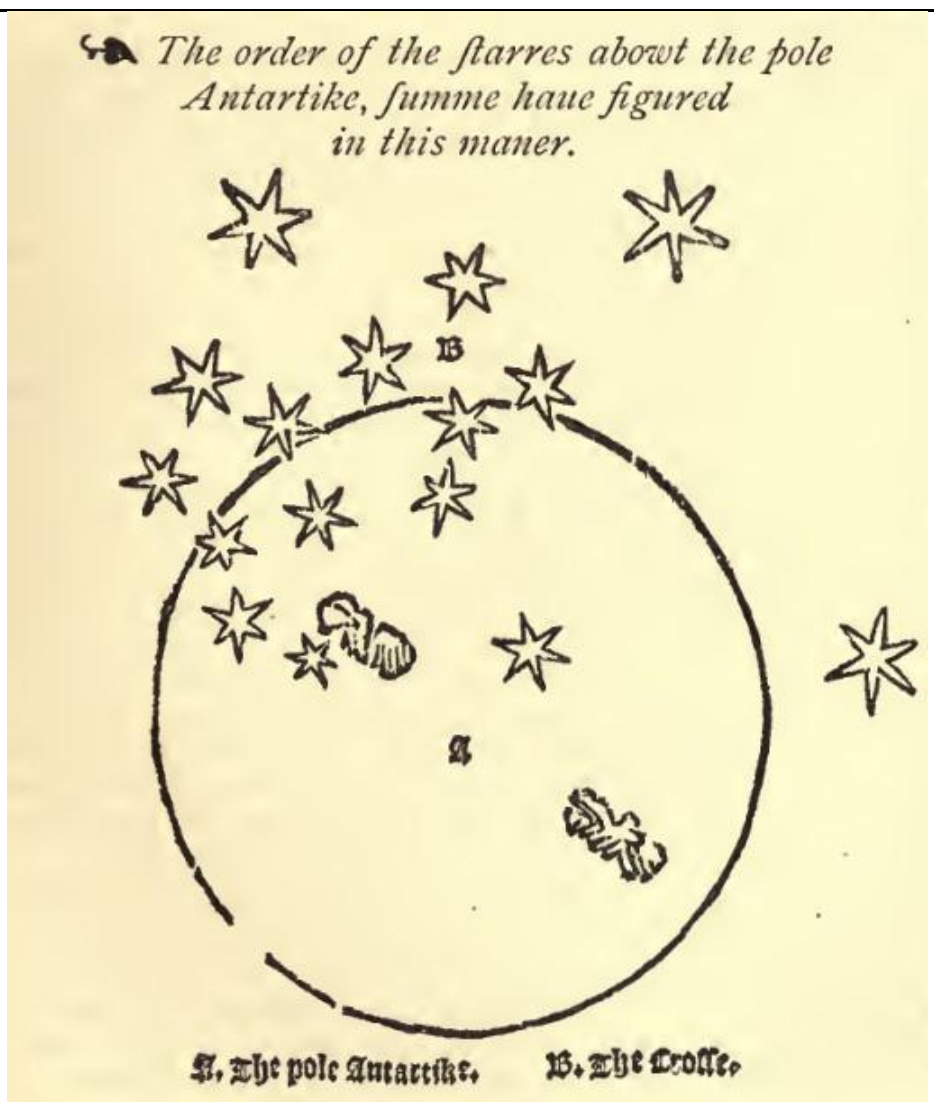 |
| Figure 3 (a) sketch of the southern sky showing Crux (top left) and below it, the Southern Triangle (Mestre João, 1500). | (b) Sketch from a letter sent by Andrea Corsali in 1515 (translated by Richard Eden). |

**Magellan's voyage and his immediate followers**

Turning now to Magellan's voyage itself, the details are known largely thanks to the lively reporting of Antonio Pigafetta, "Vicentino Cavagliere di Rodi", who accompanied the expedition as "sobresaliente" (supernumerary) to discover the world, and to seek fame (as Pigafetta admits in the introduction to his report). Although the original report, which was written for the Duke of Mantova, has disappeared, several copies or translations quickly appeared, in French, Italian, and Latin.

In an early (shortened) French edition by J. A. Fabre (Paris, circa 1526, chez Simon de Colines), which is the translation of an original sent by Pigafetta to the Duchess Louise of Savoy, regent of France at the time, one can read, after the passage about the Strait:

*Le pôle Antarctique n'est point tant étoilé comme est l'Arctique. Car on y voit plusieurs étoiles petites congrégées* (packed) *ensemble, qui sont en guise de deux nuées* (the two Clouds) *un peu séparées l'une de l'autre, et un peu offusquées* (obscured), *au milieu desquelles sont deux étoiles non trop grandes ni moult reluisantes et qui petitement se meuvent. Et ces deux étoiles sont le pôle Antarctique*.

In this description, which has no accompanying sketch, it seems that the pole is represented by two small stars in the middle of the two Clouds (possibly γ and ν Hydri), unlike in Corsali's figure. But Pigafetta was not an astronomer, merely a writer who was sensitive to the beauty of the southern sky. His description of the voyage (and hence the Clouds) has probably better survived because it is livelier than the technical reports by navigators. Yet this was not the case in the years immediately following Magellan's discovery of the Strait, when the most printed and distributed document/report was a letter by Maximilianus Transylvanus, secretary to Charles V, first printed in 1523. But it contains no mention of observations, the sky or the Clouds. D'Anghiera mentionned the Clouds in an earlier report (1516), but says nothing about the sky when describing Magellan's trip (d'Anghiera, Decade 5, 1524).

But where are the reports of the navigators and astronomers who accompanied Magellan's expedition? The chief pilot was then Andrès de San Martin, but he was killed, along with more than 20 others, on May 1st, 1521, at the diner offered by Humabon, Rajah of Cebu: his papers, which are lost today, had nevertheless survived for a while, probably confiscated by the Portuguese in Ternate when the Trinidad came back there. A later report of this voyage by Antonio de Herrera (1601) gives much more details about the astronomical observations, which come from San Martin's logbook: he mentions for instance many observations of the height of the Sun to derive latitudes, a new technique invented by the Junta dos Mathematicos in Lisbon but which required a table of declinations of the Sun for each day, presumably the Almanach Perpetuum of Abraham Zacuto (first printed in 1496, but existing in manuscript form since 1478), with simplified instructions in a Regimento de Navegar[2] . Tables of predicted eclipses (e.g. by Regiomontanus) were also used to derive the longitude, but they are obviously not frequent enough! Herrera mentions that the crew observed the solar eclipse of Oct. 11, 1520, but this is strange as, according to NASA's calculations (2006, NASA/TP-2006-214141), this eclipse was annular and should not have been observable from the bay of Santa Cruz, where the fleet was at the time, not even its penumbra. NASA's predicted time of maximum is 16:00 UT, which corresponds to about 11:26 local time, in line with Herrera's mention of 10:08 predicted at the time. And his description of the luminosity changes corresponds to what would be expected during a penombra...There was also another eclipse on April 17th the same year, total and well visible in the south of Argentina, which they could have observed from the bay of San Julian: Castanheda (1554) refers to such an observation on April 17th to estimate longitudes (with some errors...), but not in October, and also Barros does. It is not clear why Herrera puts more emphasis on the October event, probably because of selection effects in documents many years after the facts, and reminiscence of its observation in Spain (as his description alludes to). It is difficult to check today, as San Martin's documents have disappeared in the mean time. I have found also another mention of the early survival of San Martin's papers in the Real Bibliotheca del Escorial (Spain), where a compilation of earlier existing books (Bibliotheca Hispana Nova, by Nicolas Antonio Hispalensis, 1773) mentions : "Andreas de S.Martin scripsit “Del descubrimento del Estrecho de Magallanes”, teste Antonio de Herrera y Antonio Leonio". Barros (1553) also says that he had "in hand" some papers and a book written by San Martin, collected in the Moluccas by Duarte de Resende. San Martin’s observations thus percolated into reports by several authors. Other direct reports from the voyage are available also (the journal of Francisco Albo, the relation by Ginès de Mafra or the logbook of the Genovese pilot) but in none of them do we find any mention of the Clouds. Once again, this provides proof that the main interest was initially to find stars able to mark the southern celestial pole but not to map the sky, and even this approach was largely abandoned once the solar technique had been mastered.

Several expeditions to exploit the discovery of the new route were launched by Spain after the return of the Victoria (under Sebastian El Cano) in September 1522, about which direct reports exist: the one under Comendador Garcia Jofre de Loaysa, with El Cano himself and seven ships, departed in 1525 but survivors returned only in 1536[3]; the fleet send by Cortes in rescue to them, under Don Alvaro de Sayavedra, directly from New Spain (Mexico) in 1528; an ill-fated expedition under Simon de Alcazaba in 1534; etc...None of these reports mentions the name of Magellan, neither about the Clouds nor even about the Strait.

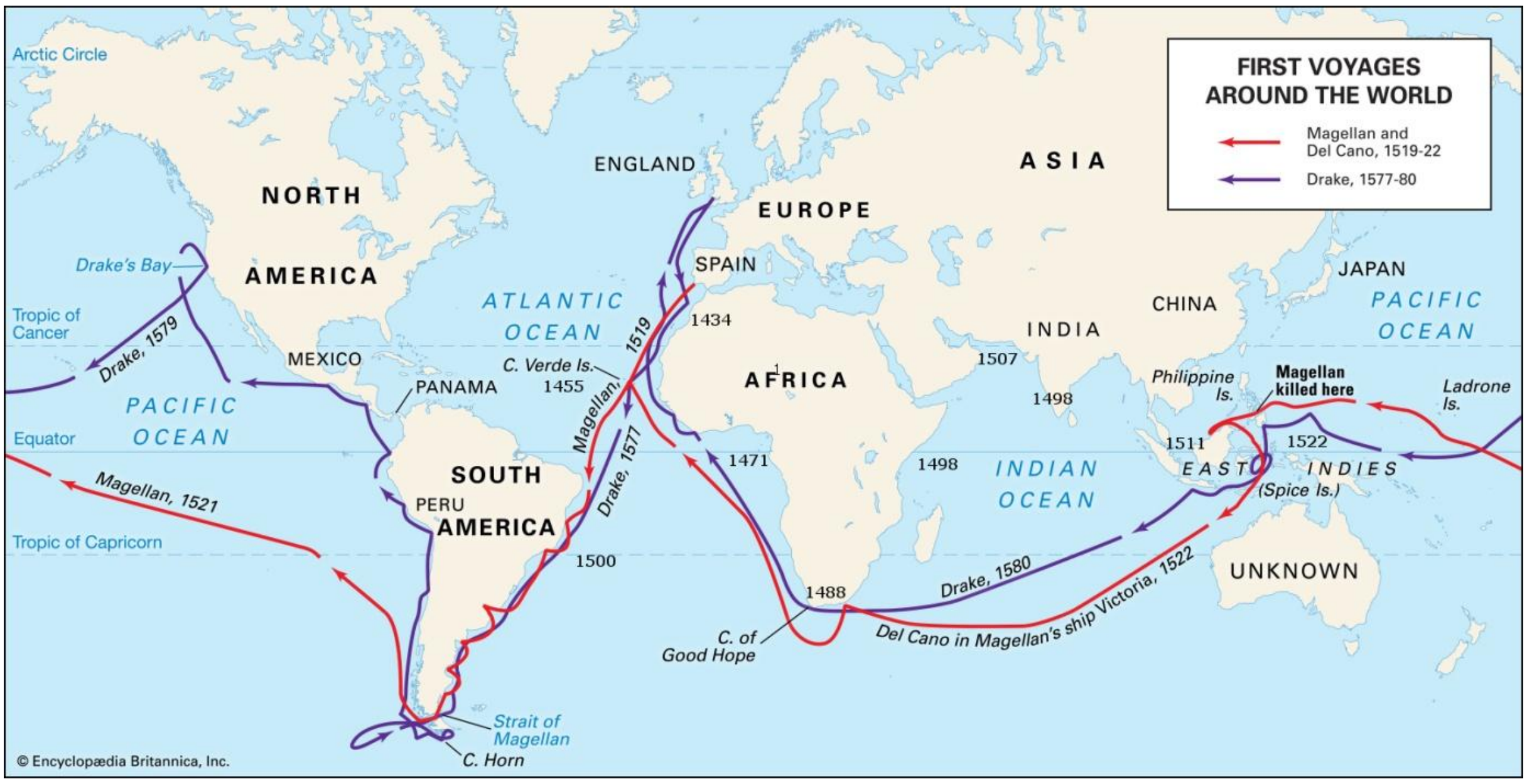

Fig. 4 Voyages of Magellan and Drake (from Encyclopedia Britannica).
I have added dates of some earlier explorations made by the Portuguese.

The fifth trip by Spaniards to the Strait was organised by the Bishop of Plasencia, who sent three ships under Alonso de Camargo, departing in August 1539. After one ship was lost in the Strait, Camargo proceeded in the third ship and went north along the west coast of Chile, becoming the first European to explore the southern part of Chile (Alonso Quintero had sailed down along the northern coast already in 1536, until the port called today Quintero, while Diego de Almagro was coming down on land). Camargo made port at Valparaiso and proceeded to Peru, where he settled down. Unfortunately no report from that ship remains. The second ship, whose name and captain have not been recorded, went back to Spain after rounding Tierra del Fuego, discovering the southern strait (now called LeMaire). A fragmented report has been discovered by Torres de Mendoza (1879) in which one finds many observations of the Sun to derive latitudes, but nothing about the sky. There is, however, a clear mention of the Strait of Magellan:

*A los 12 del dicho* (January 1540)*, surgimos junto con el cabo de las Vírgines que está en cincuenta i dos grados largos, y de allí vimos la entrada del Estrecho de Magallanes, é tiene por seña, conviene a saber: el cabo bentallado, etc*.

Later in the text, there are several mentions of the Estrecho (with capital E), showing that only about 20 years after its discovery, Magellan's name was clearly associated by sailors with the strait he had discovered[4]. In his "Cosmographie" (1544), the portuguese-french pilot and cartographer Alfonse de Saintonge ( João Afonso ?), member of what is called the Dieppe School of Cartography, says, about the Strait: "*il y a passé trois ou quatre navires de quoy estoit capitaine et pilote ung nommé Frenande (sic) de Magaillan, Portugalloys, d'où* *vient que ledict destroict est appellé Magaillan de son nom, à cause que ledict Frenande de Magaillan l'a descouvert...*".

A bit later, and unbeknownst to the Spaniards, at least at the beginning, another voyage was undertaken through the Strait and up the western coast of Chile, as far north as San Francisco. Francis Drake's voyage left Plymouth on 15 November 1577 and entered the Strait of Magellan (sic) on 20 August 1578. Again, apart from a lunar eclipse observed on 15 September that year, there is no relevant astronomical observation. I mention this here only because Houzeau (1885) claims that the name Magellan, associated with the Clouds, can be found in Hakluyt , about Drake's voyage of 1578. Besides Hakluyt's (1589) compilation, all the documents relating to that voyage, including a detailed report by the pilot Nuño da Sylva himself, can be found in Nuttall (1914) — but I could not find any mention of the Clouds! It is anyway quite unlikely that a portuguese pilot would call them differently than "Clouds of the Cape", as they used to travel to the Indies only via the eastern route around the Cape of Good Hope.

It therefore seems clear that by then the name of Magellan was being associated by navigators with the Strait, but not necessarily with the Clouds. This is also indicated by the report of one of the most successful expeditions to the Strait, conducted by the brothers Nodal in 1618–1619, who made the journey in less than 10 months and, remarkably, with no losses. It was said to have been organised to recognise the "*new strait of São Vicente and the one of Magellan*", because "*it had become difficult owing to the number of years during which notice of the navigation had been lost*," as Sir Clements Markham's translation (Markham, 1911) has it[5]. This report mentions some celestial observations (including the sight of the great Comet of 1618) , but there is no mention of the Southern Cross or of the Clouds, as if these were already common knowledge. Clearly, by this point the Sun was the preferred way of determining latitudes, together with tables in the Hidrografia Nautica of Cespedes (1606).

## First mappings of the Southern Sky

The Spaniards' interest in the Spice Islands progressively declined, as they were merely active in getting silver and gold back from Péru and had sold their "rights" on the Moluccas to Portugal (1529, Treaty of Saragossa). The Dutch launched their first expedition, known as the "*eerste Schipvaart*", under Cornelis de Houtman, who left Holland for the Indies on 2d of April 1595. On board were his brother Frederick and Pieter Dirkszoon Keyser (latinised as Petrus Theodori). Keyser had been trained by the cartographer Petrus Plancius, who instructed him to make a record of the southern sky for his celestial globes. This is the first known systematic measurement of the southern sky, probably mostly carried out when the fleet stopped for several months in Madagascar. Unfortunately, Keyser died in Indonesia (reached only in June 1596), but the measurements were presumably brought back by Frederick de Houtman in August 1597. During a second voyage with his brother, Frederick complemented those observations further, returning in July 1602. None of Keyser's notes survived, but de Houtman published measurements, which were long hidden as an appendix to his dictionary and grammar of the Malayan and Malagasy languages (Houtman, 1603), later reprinted as a catalogue of southern circumpolar stars in a French translation by Aristide Marre (1881). They have clearly been used on various celestial globes by Hondius (1598; 1600; 1601), Willem Blaeu (1602) and ultimately by Bayer in his Uranometria (Augsburg, 1603) where he explicitly mentions Petrus Theodori as the originator of his 12 new constellations (although it is actually Plancius, not Theodori, who is likely behind these twelve constellations of the southern sky, still in use today with small modifications). Interestingly, no mention of the Clouds is made in Houtman's catalogue but they do appear on Hondius's earlier globes. For instance, of Dorado (where the LMC would be) Houtman (1603) says only:

*Den Dorado heft 4.Sterren*,

and in

*Den Indiaenschen Exster, op Indies Lang ghenaemt* (our Tucana, close to the SMC),

he mentions only six stars. But Bayer's map of the southern pole (his plate 49) indicates both Nubecula Minor and Nubecula Major (without mention of Magellan), Latin being the usual working language at the time.

A detailed analysis of these early mapping attempts can be found in Dekker (1987). Houzeau (1885) claims that Theodori returned in 1597, bringing back

*une énumération des Constellations du Sud en douze astérismes antarctiques*…,

one of which was "*le Nuage*" (The Cloud). Houzeau initially counts 12 new constellations including The Cloud, but then ends up with 21, eight of which were already listed by Ptolemeus: his total now amounted to 13 (21 – 8), including two new ones (Crux and Musca) but excluding The Cloud ! It is clear that Houzeau was misled about Theodori by the comment in Bayer's Uranometria: "Duodecim haec novas exhibet Schediographias...quas ..., Petrus Theodori nauclerus peritissimus, novissimo annotatas,...divulgavit."). But it is less clear where his idea of the constellation *le Nuage* comes from: maybe he was influenced by the presentation of the Clouds in the earlier globes of Hondius (see Figure 5) or Blaeu. There are unfortunately many approximate statements in this paper. As for globes, see a detailed description of some of them in van der Krogt (1993) or at the National Maritime Museum in Greenwich, but these are all references to the Nubeculae, not to the name Magellan.

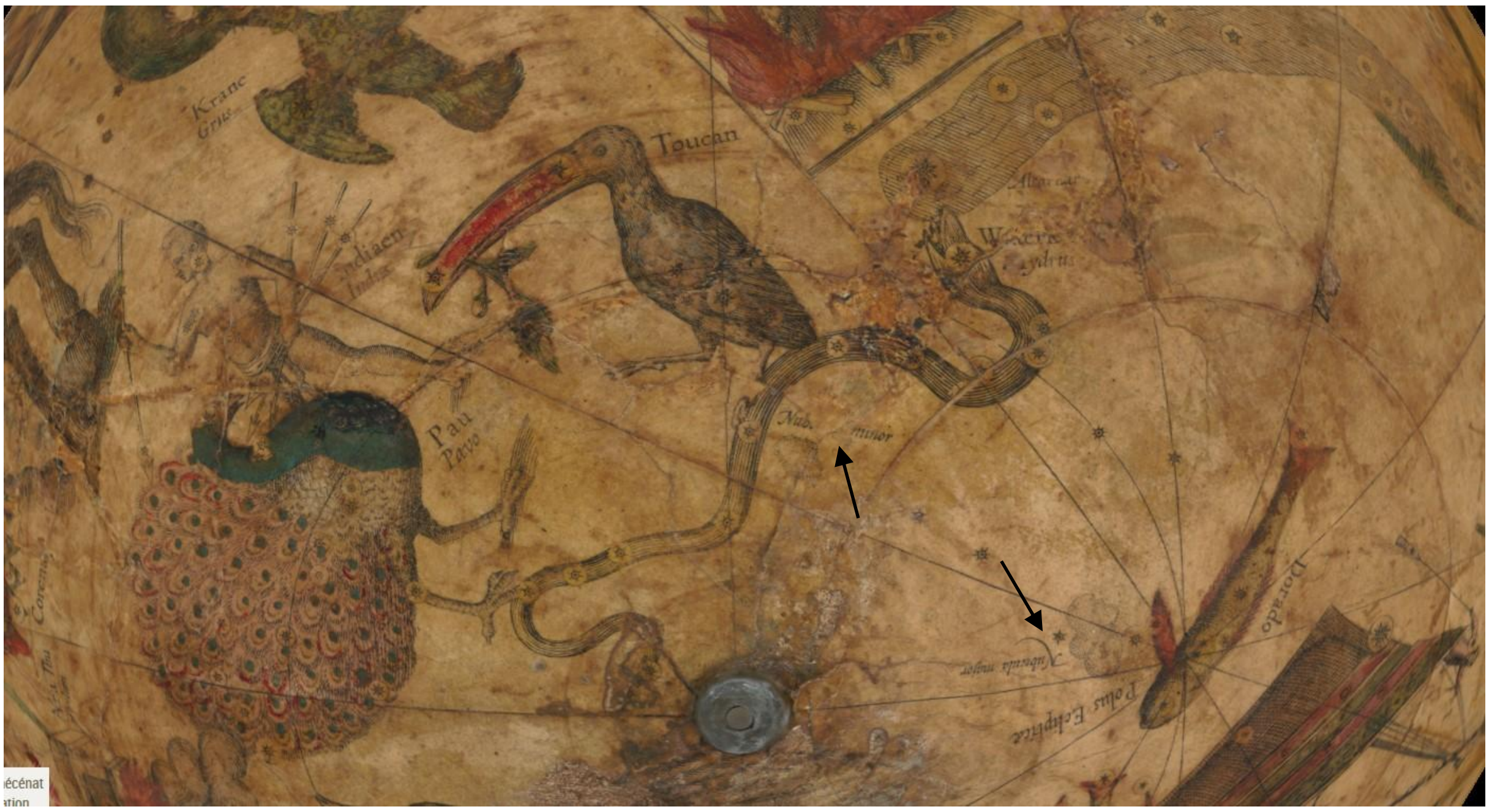

Fig. 5   The Toucan (Hondius globe, 1600, BNF Paris)

Below Tucana is Nubecula Minor; and Nubecula Major is below Dorado, in the lower right corner (arrows)

It is clear from these early records that neither the Clouds nor the Southern Cross were of any real interest to the sailors anymore, rather they were merely a curiosity in the sky. The Portuguese, who used the eastern route via the Cape of Good Hope (initially called Cabo Tormentoso in 1488 by Bartolomeu Dias...but Good Hope was deemed to be a better selling name by the King of Portugal), of course knew them early and occasionally mentioned them as the Clouds of the Cape. The development of celestial cartography proceeded and usually included the Clouds, but with their Latin names, Nubecula Minor and Nubecula Major (see, for example, the maps in Bayer's atlas third edition of 1661; or in the atlases of Cellarius (Amsterdam 1660); Pardies (Paris 1675); Flamsteed (London 1725);  Ottens (Amsterdam 1729) or Doppelmayr (Nürnberg, 1742). Schiller (Augsburg, 1627), in his curious biblical map, includes Tucana, Hydrus and Nubecula Minor in his idiosynchratic constellation of St Raphaël. Bode, in his Uranographia (Berlin, 1801), where the names of 30 Dor or 47 Tuc come from,  uses both German and French, but does not mention Magellan.

Two major efforts to get better celestial cartography in the south were undertaken after Plancius. The first was by Edmond Halley from the island of St Helena, where he had been sent by John Flamsteed the first English Astronomer Royal; he observed during 1677 and published the Catalogus Stellarum Australium (Halley, 1679). His work is known because of the inclusion of a new (fleeting) constellation, Robur Carolinum, to please King Charles II. He mentions in an appendix that:

*Proxima, ex iis quas observavi, est in Cauda Apodis, in distancia paulo ultra 8 graduum, Duae Nubeculae, quae a Nautis Nebulae Magellanicae appellantur, exacte referent Galaxiae albedinem & Telescopio inspectae, hinc inde Nebulas Parvas & exiguas Stellas ostendunt…*

It appears it was clear to him that the name Magellanic Clouds was used only by sailors, and that those Clouds resemble the Milky Way. The only other mention in the catalogue itself comes in the final list of bright stars "*in Usum Navigantium*", useful for navigation, where the first in the list is "*quae adjacent Nubeculae minori*", close to the Small Cloud. The Clouds also appear as Nubeculae in his Planisphère of 1678.

The second effort was by Nicolas de La Caille who observed at the Cape of Good Hope between August 1751 and July 1752. He published a catalogue of southern nebulae (or, more precisely, of nebulous stars) in 1755, in which the Clouds are not listed, probably because they were too large for his purpose (47 Tuc and 30 Dor are mentioned). In the introduction he says that he observed the brightest regions of the Milky Way several times with a 14-foot (4.27-m) focal length telescope, and compared them to the two Clouds

*"qu'on appelle communément les nuées de Magellan et que les Hollandais et les Danois appelent les nuées du Cap* (his underlining). *On voit évidemment que ces parties blanches du ciel se ressemblent si parfaitement, qu'on peut croire, sans trop donner aux conjectures, qu'elles sont de même nature,..."*

He adds that these Clouds seem to be detached parts of the Milky Way, and that it is not clear that their whiteness would be caused, as is usually believed (sic), by clusters of small stars more tied together than in other parts of the sky, as he could not resolve them with his telescope. About the names, he also says:

*D'ailleurs, la plupart des Navigateurs appellent nuages du Cap , ce que nous appelons nuées de Magellan, ou le grand & le petit nuage*.

In the celestial map completing his first publication (the catalogue of 1752, see Figure 6) the two Clouds do indeed appear under the names "*le Petit, et le Grand Nuage*". But his words, "*que nous appelons nuées de Magellan*", seem to indicate that around that time the association of the Clouds with Magellan was already spreading beyond the nautical community, even if the scientific term was still simply Nubeculae or les Nuages.

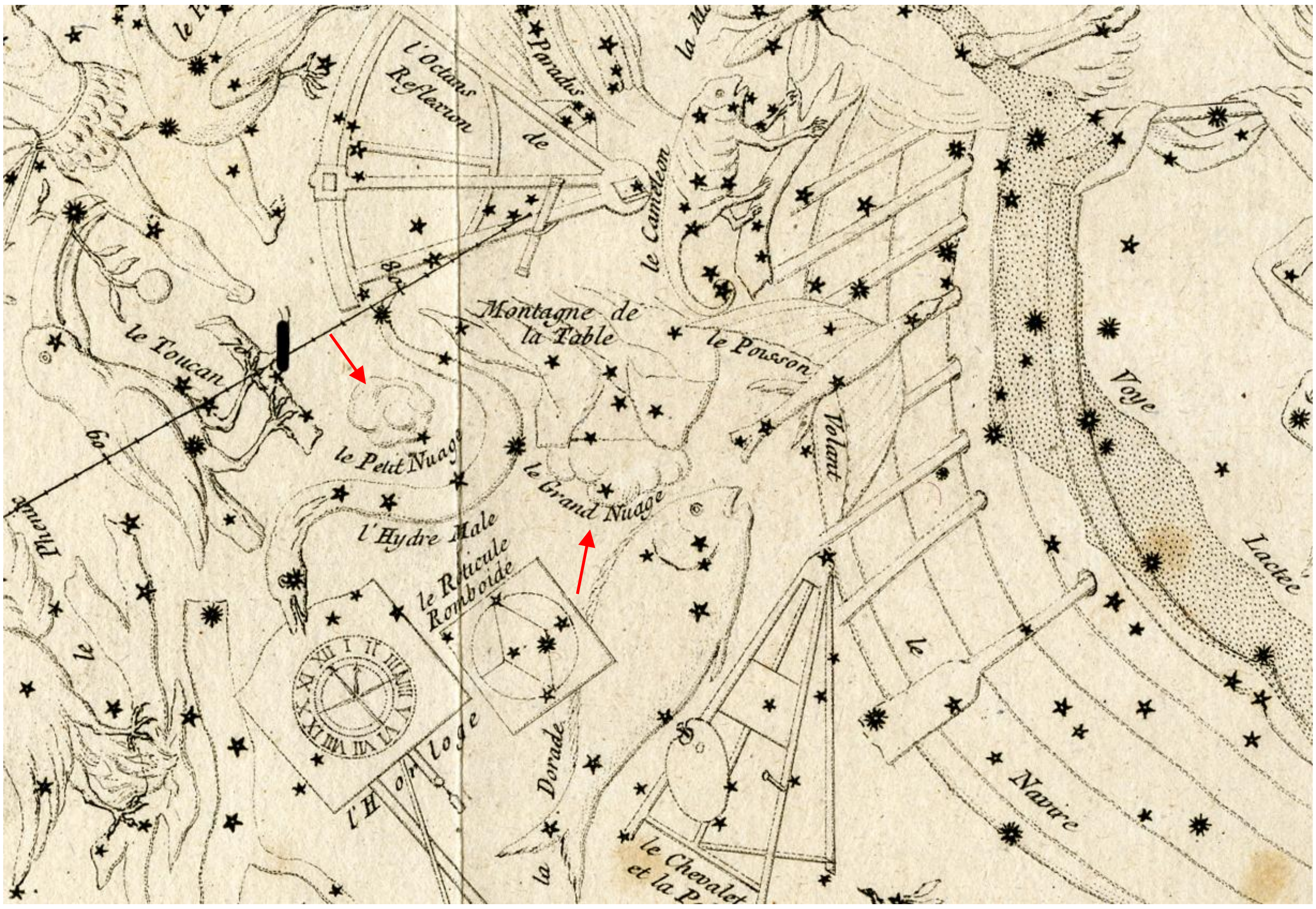

Fig. 6  Enlargement of La Caille's map of 1752, showing the 2 Clouds (red arrows; Paris Observatory)

When James Dunlop described his observations of the southern sky (Dunlop, 1828) "made at my house" in Paramatta, near the Brisbane Observatory in Australia, he reported that he made: "*very correct drawings of the Nebulae major and minor...with an excellent 9-feet reflecting telescope*" but there is not a single mention of Magellan. He gives detailed sketches of both nebulae, with the positions of many stars and smaller nebulae within (Fig.7).

When Rümker later published his catalogue (Rümker, 1832) based on observations from the Observatory of Paramatta, Australia, he states that "*the Nubeculae major and minor of La Caille are two fragments of the via lactea and distinguish themselves by nothing from any other parts of it requiring, with the exception of two nebulae, no powerfull telescope to be dissolved in well-defined Star's*" — again, no mention of Magellan.

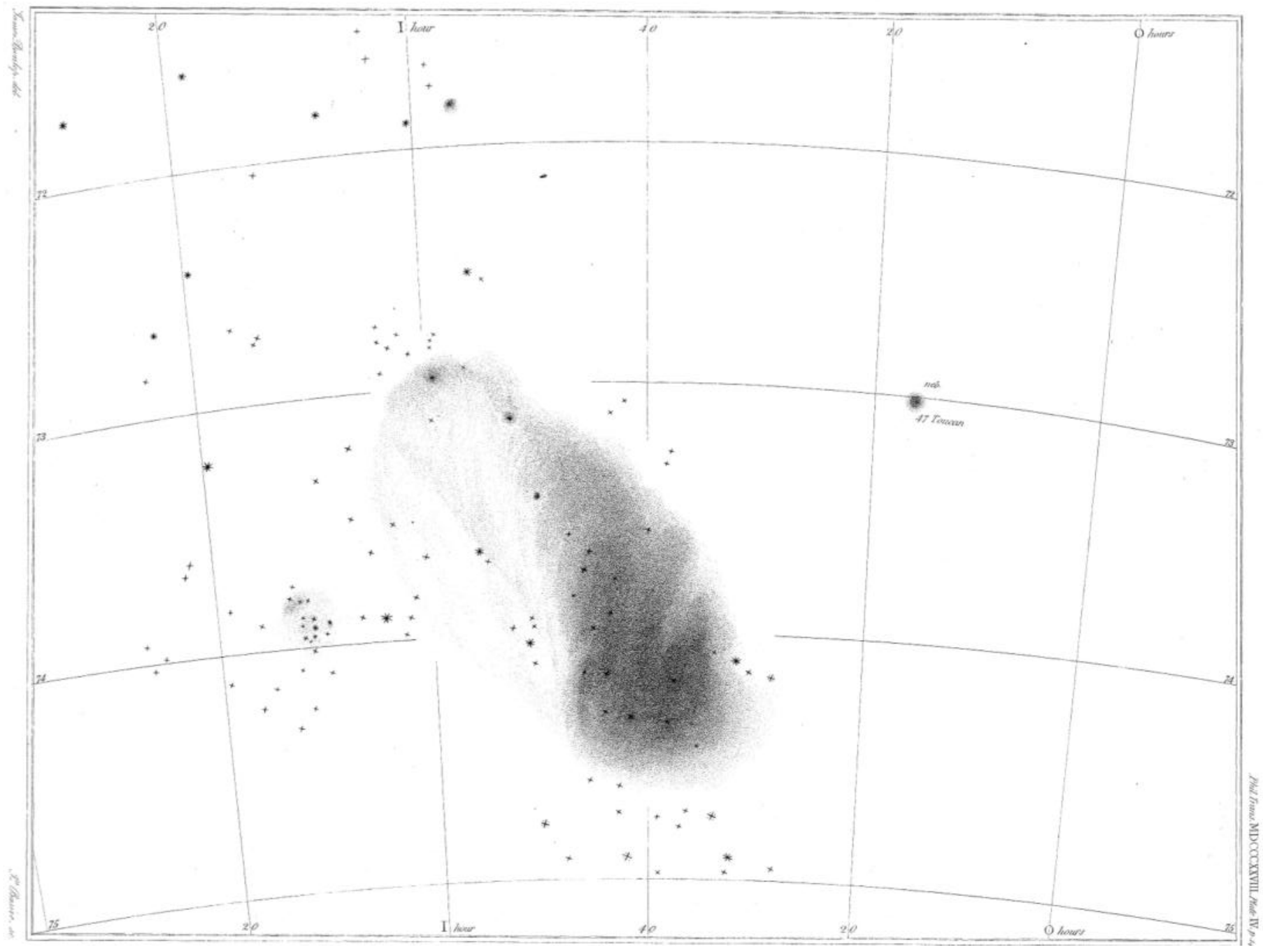

Fig.7  The SMC as drawn by Dunlop (1828)

John Herschel (1847), in a section however entitled "On the two nubeculae or Magellanic Clouds", talks only about Nubeculae. Only in the accompanying figure (Fig. 8), where he gives his visual observations (figure where it is however hard to recognise the Clouds),  does he mention again , "*The two Magellanic Clouds as seen with the naked Eye*". So he uses both the scientific denomination and the more public name, and Herschel seems to have been the first one  to use the name Magellanic Clouds in a scientific publication.

Later, in the 20th century, in the first edition (1910) of the well-known Norton's Star Atlas, it is stated that, "*the Magellanic Clouds or Nubecula Major and Nubecula Minor appear to the naked eye like detached portions of the Milky Way, and are a marvelous sight in the telescope*" as if their names were obvious, but without any further note on Magellan (nor on their nature). So, although no precise date can be given, it seems that by the late 19th century, the term Nubecula was still being used in scientific exchanges, but the term Magellanic Clouds was progressively passing from nautical circles to the public and scientific spheres, finally replacing Nubeculae only once scientists abandoned Latin.

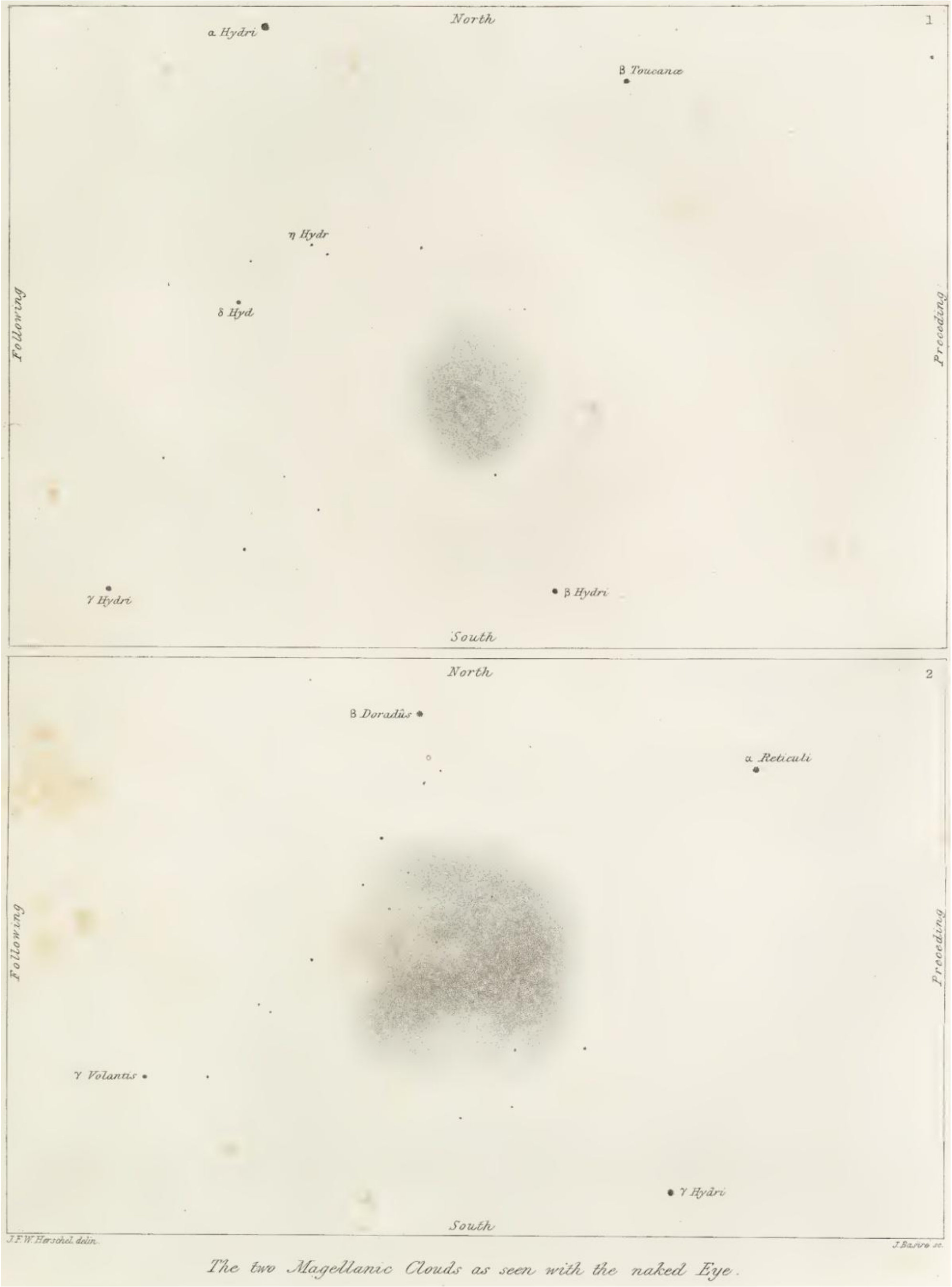

*Fig. 8  "The two Magellanic Clouds as seen with the naked Eye" (Herschel, London 1847)*

To conclude about Magellan, one may turn to words from the introduction to Drake's famous voyage ( Ed. 1628, dedicated to the Truly Noble Robert Earle of Warwicke):

*Fame and envie are both needlesse to the dead because unknown, sometimes dangerous to the living when too well knowne: reason enough that I rather chuse to say nothing, then too little, in the praise of the deceased Author.*

It is very difficult today to observe "at my house", as Dunlop did, but astronomers travelling to Chile should keep in mind how hard it was to reach that country in those early days. Today, we do it in 14 hours, instead of 14 months.

**Notes**

1) It is now admitted that Cabral was not the first one to discover Brasil for Portugal, end of April 1500. Probably Duarte Pacheco reached it already in 1498, as he mentioned himself (see his Esmeraldo (1508) and also Barros (1552)), but no official report was given at the time (perhaps due to secrecy…but as today, no publication = no credit). But Vespucci did, in his spanish voyage with Alonso de Hojeda and Juan de la Cosa (his second voyage): he reached Brasil, at its NE coast, in June 1499. Then Vicente Yañez Pinzon did in January 1500, and Diego de Lepe in April the same year, all three for the King of Spain. Not much publicity was given at the time, probably because they were traveling in the zone reserved to Portugal...

2) Readers might be interested to know that the oldest surviving manual for navigation, the Portuguese *Regimento do Estrolabio*, which probably dates from 1509, sits not in Evora, but in Munich (ex Royal Library), (see “Seltenheiten aus Süddeutschen Bibliotheken” (E. Freys, München, 1912), and Bensaúde, 1912).

3) It has been suggested (Roger Hervé, Paris, 1982) that the Santo Lesme, which lost the fleet at the exit of the strait, in making her way alone to the Moluccas had probably made a southern route and discovered New Zealand, to finish as a wreck on the southern coast of Australia (the Mahogany ship, still to be recovered and explored). Another story to investigate, together with the probable early discovery of Australia by the Portuguese, directly from the Moluccas.

4) There is an earlier indication that some sailors had already associated the name of Magellan to the Straight during the discovery voyage itself, but this did not seem to have been noted immediately. It comes in the "Relation of an anonymous portuguese who accompagnied Duarte Barbosa" (first published in italian by Ramusio in 1554), which says: " *...navegámos ao longo da costa 378 milhas...donde nos encontrámos num estreito, ao qual pusemos nome estreito da Victoria, porque a nau Victoria foi a primeira que o viu; alguns lhe chamaram o estreito de Magalhães, porque o nosso capitão se chamava Fernando de Magalhães. A boca deste estreito está em 53 graus e meio* " (see Garcia, 2007, for details).

5) This expedition was in fact launched in reaction to an attempt by the Dutch to reach the Moluccas by the western route (the “Spanish route”): a vessel sponsored by Isaac LeMaire was sent under the orders of Cornelisz Schouten in June 1615, discovered a new strait south of the Tierra del Fuego (later named after LeMaire, although Camargo's fleet, perhaps one of Loaysa's ships and probably Drake also, had already discovered it earlier) and the most southern land in January 1616, which they named Cape Hoorn, after the home town of the Captain ( Drake had named it Cape Elisabeth).

Ackowledgements

Many of the texts have not reached us in their original form, but rather in copies or Latin, French (or, later, English) translations. Whenever possible, I have presented text in the original language (except Arabic), to also reflect the diversity of languages in use in the European community. Concerning travel and the discovery of the southern hemisphere, some of the compilations mentioned above — to which the reader may refer for more details — such as de Barros (1552), d'Anghiera/Eden (1516/1555), Ramusio (1550), Herrera (1601) or Fernández de Navarrete (1837), are available online; some travel reports have been translated into English by the Hakluyt Society (which honours the work of Richard Hakluyt himself [1553–1616]). Regarding Magellan's travel, most available documents have been assembled in the excellent, recent book by de Castro (2010). Many detailed reports of Portuguese travels (by da Mosto, Vespucci, Gama, etc.) have also been published by Edts Chandeigne (Paris).

This work is of course not the first one dealing with the discovery and description of the southern sky, but concentrates on a largely unexplored topic, namely the reporting and naming of the Clouds. While conducted independently, it has also benefited, in addition to the references and compilations quoted above, from prestigious predecessors. To mention at least the most important ones: the Kosmos from Alexander von Humboldt, the 3rd volume of which appeared in 1851; the report by E. Halley, which appeared in 1679 in Latin in London, and the same year in Paris translated into French; and the reports of La Caille, which appeared in Paris in the Mémoires de l' Académie des Sciences in 1752 (p.586) and 1755 (p.194). See also the IAUS 190 in 1998 where Paul Hodge dealt with that topic (IAUS190, Edts Y.H.Chu et al., ASP 1999, p.3). A biography of La Caille has been published in 2013 by Ian Glass (Oxford Univ. Press), translated into French by J. Lequeux (2013, Paris: EdP Sciences).

It is a pleasure to thank Hakim Atek and Carlos Carvalho (IAP), Rodolfo Barba (Univ. La Serena), Beatriz Barbuy (Univ. São Paulo), Michel Chandeigne (Librairie Portugaise, Paris), Muriel Dennefeld (Science Po Paris), Carl Ehrig-Eggert (Univ. Frankfurt*),* Emilie Kaftan (Paris Observatory), Somaya Saad and Ahmed Shokry (Cairo Observatory), Mira Veron (OHP), Hans Zinnecker (ESO-Chile) and Louise Devoy and Scarlet Faro (National Maritime Museum, Greenwich) for useful exchanges and copies of some documents.